\documentclass[showpacs,preprint,amsmath,superscriptaddress,prl]{revtex4} 
\usepackage[dvips]{graphicx}
\usepackage{epsfig}
\usepackage{color}
\usepackage{subfigure}
\usepackage{soul}

\usepackage{amsmath}
\usepackage{amssymb}

\begin{document} 
\title{
From Boltzmann to Zipf
through Shannon and Jaynes
} 
\author{\'Alvaro Corral}
\affiliation{Centre de Recerca Matem\`atica,
Edifici C, Campus Bellaterra,
E-08193 Barcelona, Spain.
} 
\affiliation{Departament de Matem\`atiques,
Facultat de Ci\`encies,
Universitat Aut\`onoma de Barcelona,
E-08193 Barcelona, Spain}
\affiliation{Barcelona Graduate School of Mathematics, Edifici C, Campus Bellaterra, E-08193 Barcelona, Spain}
\affiliation{Complexity Science Hub Vienna, Josefstädter Stra$\beta$e 39, 1080 Vienna, Austria}
\author{Montserrat Garc\'{\i}a del Muro}
\affiliation{Departament de F\'{\i}sica de la Mat\`eria Condensada,
Universitat de Barcelona, Mart\'{\i} i Franqu\`es 1, E-08028 Barcelona, Spain}
\affiliation{IN2UB, Universitat de Barcelona, Mart\'{\i} i Franqu\`es 1, E-08028 Barcelona, Spain}
%
%
\begin{abstract} 
The word-frequency distribution provides the fundamental building blocks 
that generate discourse in language. 
It is well known, from empirical evidence, 
that the word-frequency distribution of almost any text 
is described by Zipf's law, at least approximately. 
Following Stephens and Bialek [Phys. Rev. E 81, 066119, 2010], 
we interpret the frequency of any word as arising from 
the interaction potential between its constituent letters. 
Indeed, Jaynes' maximum-entropy principle, 
with the constrains given by every empirical two-letter marginal distribution, 
leads to a Boltzmann distribution for word probabilities, 
with an energy-like function given by the sum of all pairwise (two-letter) potentials. 
The improved iterative-scaling algorithm allows us finding the potentials from the empirical two-letter marginals. 
Appling this formalism to words with up to six letters
from the English subset of the recently created 
Standardized Project Gutenberg Corpus,
we find that the model is able to reproduce Zipf's law,
but with some limitations:
the general Zipf's power-law regime is obtained, 
but the probability of individual words shows considerable scattering.
In this way, a pure statistical-physics framework is used to describe
the probabilities of words.
As a by-product, we find that both the empirical two-letter marginal distributions
and the interaction-potential distributions follow well-defined statistical laws.
\end{abstract} 

\date{\today}

\maketitle

\section{Introduction}

Zipf's law is a pattern that emerges in many complex systems 
composed by individual elements that can be grouped into different
classes or types \cite{Li02}.
It has been reported in demography,
with citizens linked to the city or village where they live \cite{Malevergne_Sornette_umpu};
in sociology, with believers gathering into religions \cite{Clauset};
in economy, with 
employees hired by companies
\cite{Axtell};
and also in ecology \cite{Pueyo,Camacho_sole},
communications \cite{Adamic_Huberman,Clauset},
cell biology \cite{Furusawa2003},
and even music \cite{Zanette_music,Haro,Serra_scirep}.
In all these cases, the size of the groups in terms 
of the number of its constituent elements 
shows extremely large variability, 
more or less well described in some range of sizes by a power-law distribution 
with an exponent close to two (for the probability mass function; 
this turns out to be an exponent close to one for the complementary
cumulative distribution).
Of particular interest is Zipf's law in linguistics \cite{Baayen,Baroni2009,Zanette_book,Piantadosi,Altmann_Gerlach,Moreno_Sanchez}, 
for which individual elements are word tokens 
(i.e., word occurrences in a text), 
and classes or groups are the words themselves (word types).
In this way, the ``size'' of a word type is given by the number of tokens
of it that appear in the text under study (in other words, the absolute frequency of the word).

There have been many attempts to provide a mechanism for this curious law
\cite{Mitz,Newman_05,Loreto_urn}.
With text generation in mind, we can mention monkey typing, 
also called intermittent silence \cite{Miller_monkey,Ferrer-i-Cancho_2010},
the least effort principle \cite{Ferrer2002a,Prokopenko,Dickman_Moloney_Altmann},
sample-space reduction \cite{Corominas_dice}, 
and codification optimization \cite{Ferrer_cancho_compression}.
More general mechanistic models for Zipf's law are preferential attachment \cite{Simon,Cattuto,Zanette_2005,Gerlach_Altmann},
birth-and-death processes \cite{Saichev_Sornette_Zipf},
variations of Polya urns \cite{Tria}
and random walks on networks \cite{Perkins}.
The existence of so-many models and explanations is a clear indication 
of the controversial origin of the law.
Further, there have been also important attempts to explain not only Zipf's law
but any sort of power-law distribution in nature \cite{Bak_book,Sethna_nature,Sornette_critical_book,Watkins_25years}.

A different approach is provided by the maximum-entropy principle.
In statistical physics it is well known that a closed system in equilibrium with 
a thermal bath displays fluctuations in its energy but keeping a constant
mean energy. 
As Jaynes showed \cite{Jaynes57},
the maximization of the Shannon entropy with the constrain that the mean
energy is fixed yields the Boltzmann factor, 
which states that the probability of any microstate has to be an
exponential function of its energy 
(note that this does not mean that the distribution of energy is exponential, 
as the number of microstates as a function of the energy
is not necessarily constant).

Therefore, some authors have looked for an analogous of the Boltzmann
factor for power laws.
For example, one can easily obtain a power law not imposing a constant
(arithmetic) mean but a constant geometric mean \cite{Nieves}
(assuming also a degeneracy that is constant with respect the energy).
Also, fixing both the arithmetic and the geometric mean 
leads to a power law with an exponential tail
\cite{Main_information}.
Nevertheless, the physical meaning of these constraints is difficult
to justify.
More recently, Peterson et al. \cite{Peterson_Dill}
have proposed a concrete non-extensive energy function
that leads to power-law tails of sizes when maximizing the Shannon entropy.
The main idea is that the probability is exponential with the energy, 
but the energy is logarithmic with size, resulting in an overall power law for sizes.
Other authors have found the use of the Shannon entropy inadequate, 
due to its close connection with exponential distributions, 
and have generalized the very entropy concept,
yielding non-extensive entropies such as the Havrda-Charv\'at entropies
\cite{Havrda_Charvat}, 
also called Tsallis entropies \cite{Tsallis_bjp},
and the Hanel-Thurner entropies \cite{Hanel_Thurner,Hanel_Thurner_2}.

Here we will follow the distinct approach of Stephens and Bialek \cite{Stephens_Bialek}. 
As Peterson et al. \cite{Peterson_Dill}, 
these authors consider maximization of the plain
Shannon entropy, but in contrast to them, 
no functional form is proposed ``a priori'' for the energy.
Instead, the constrains are provided by the empirical two-body marginal distributions.
The framework is that of word occurrence in texts, 
and words are considered as composed by letters that interact in pairs.
The interaction potentials are provided in a natural way by the Lagrange multipliers
obtained in the maximization of entropy under the empirical values of the constrains.

Stephens and Bialek \cite{Stephens_Bialek} only considered four-letter English words and performed 
a visual comparison with the empirical frequencies of words. 
We will considerably extend their results by analyzing words of any length from 1 to 6 letters, and will undertake a quantitative statistical analysis of the fulfillment of Zipf's law.
We will pay special attention to the values of the interaction potentials.
The main conclusion is that two-body (two-letter) pairwise interactions
are able to reproduce a power-law regime for the probabilities of words 
(which is the hallmark of Zipf's law),
but with considerable scatter of the concrete values of the probabilities.

In the next section we review the maximum-entropy formalism and its
application  to pairwise interaction of letters in words, using the useful concept of feature functions.
Next, we describe the empirical data we use and the results, 
including the empirical pairwise marginals (which are the input of the procedure)
and the resulting pairwise potentials 
(which are the output from which the theoretical word distribution is built).
The Zipfian character of the theoretical word distribution 
as well as its correspondence with the empirical distribution is evaluated.
In the final section we discuss limitations and extensions of this work.

\section{Maximum entropy and pairwise interactions}

``Information theory provides a constructive criterion for setting up probability distributions on the basis of partial knowledge,'' 
which leads to a special type of statistical inference.
This is the key idea of Jaynes' maximum-entropy principle \cite{Jaynes57}.
The recipe can be summarized as: 
use that probability distribution which has maximum Shannon entropy
subject to whatever is known.
In other words, 
everything should be made as random as possible, but not more
\cite{Broderick}
\footnote{E. G. Altmann has made us notice that Jaynes, being close
to be a Bayesian, would not have totally agreed with the identification of entropy with randomness, and would have prefer the use of ``ignorance''.
So, we could write instead: 
we should be as ignorant as possible, but not more.
}.

Let us consider words in texts.
Labelling each word type by $j$, 
with $j=1,2, \dots V$, and $V$ the size of the vocabulary
(the total number of word types), the Shannon entropy is
$$
S=-\sum_{j=1}^V P_j \ln P_j,
$$ 
where $P_j$ is the probability of occurrence of word type $j$.
Note that as we use natural logarithms, 
the entropy is not measured in bits but in nats, in principle.
In order to maximize the entropy under a series of constrains
one uses the method of Lagrange multipliers, 
where one finds the solution of
\begin{equation}
\frac {\partial\mathcal{L}}{\partial P_j} = -\ln P_j -1
-\alpha \frac {\partial}{\partial P_j} (\mbox{constrain 1})
-\beta \frac {\partial}{\partial P_j} (\mbox{constrain 2}) -\dots
= 0, 
\label{EqLagrange}
\end{equation}
for all $j$,
with $\alpha$, $\beta$, etc. 
the Lagrange multipliers associated to constrain 1, constrain 2, etc.,
and $\mathcal{L}=S 
- \alpha \times (\mbox{constrain 1})
- \beta  \times (\mbox{constrain 2}) - \dots$
the Lagrangian function.

One can see that the maximum-entropy method yields intuitive solutions
in very simple cases.
For example, if no contrains are provided one obtains the equiprobability case, 
$P_j^{\mu c}=1/V$ (as there is in fact one implicit constrain: normalization;
$\mu c$ stands from microcanonical, in analogy with statistical physics).
If there are no other constrains it is clear one cannot escape
this ``rudimentary'' solution. 
%
If, instead, one uses all empirical values as constrains, one gets the same one puts, 
with a solution $P_j^{full}=\rho(j)$, 
with $\rho(j)$ the empirical probability of occurrence of word $j$
(i.e., the relative frequency of $j$).
So, the full data is the solution, 
which is of little  practical interest, 
as this model lacks generalization and does not bring any understanding.
More interestingly,
when the mean value of the energy is used as a constrain
(as it happens in thermodynamics
for closed systems in thermal equilibrium with a bath),
the solution is given by the Boltzmann distribution
\begin{equation}
P_j^{can} = \frac{e^{-\beta E_j }} Z,
\label{EqBoltzmann}
\end{equation}
with the term $can$ coming from the analogy with the canonical ensemble
and with $Z=\sum_j e^{-\beta E_j }$.
Needless to say, we have no idea yet what the energy $E_j$ of a word is.

\subsection{Feature functions and marginal probabilities}

At this point it becomes useful to introduce the feature functions
\cite{Berger_etal}.
Given a feature $i$, the feature function $f_i(j)$
 is a function that for each word $j$ takes the values
$$
f_i(j) =
\left\{\begin{array} {ll}
1 & \mbox{if the word $j$ contains feature $i$}\\
0 & \mbox{if not}\\
\end{array} \right.
$$
For example, let us consider the feature 
$i= \{$ letter {\tt c} is in position $1 \}$, 
summarized as $i=1c$;
then $f_{1{\tt c}}(\mbox{\tt cat})= 1$
and $f_{1{\tt c}}(\mbox{\tt mice})= 0$,
as {\tt c} is the first letter in {\tt cat}  
but not in {\tt mice}
(let us mention that, for us, capital and lower-case letters are considered the same letter).

Considering $m$ features, each one yielding a constrain
for its expected value, we have
\begin{equation}
\langle f_i\rangle =
\sum_{j=1}^V 
P_j f_i(j) = F_i
\label{EqConstrain0} 
\end{equation}
for $i=1,2,\dots m$, with $F_i$ the empirical value of feature $i$.
Note that $P_j$ and $\langle f_i\rangle$ are unknown, 
whereas $F_i$ should not.
With these $m$ constrains,
the method of Lagrange multipliers [Eq. (\ref{EqLagrange})] leads to 
$$
\frac {\partial\mathcal{L}}{\partial P_j} = -\ln P_j -1 {+}
\sum_{i=1}^m \lambda_i  f_i(j)= 0,
$$
where $\lambda_i$ are now the Lagrange multipliers
(we have in fact inverted their sign with respect the previous examples, 
for convenience). The solution is
\begin{equation}
P_j = \exp \left({-1 + \sum_{i=1}^m \lambda_i  f_i(j)} \right) 
\label{solution}
\end{equation}
$$=
\exp\left(-1+\sum\lambda\mbox{'s of features of word }j\right).
$$

In contrast with the previous simplistic models, 
we are now able to deal with the inner structure of words, 
as composed by letters, i.e., $j= \{\ell_1,\ell_2,\dots\}$
and $P_j=P(\ell_1 \ell_2\dots),$
with $\ell_1$ the letter at first position of word $j$ and so on.
If we consider that the features describe the individual letters of a word, 
for example, for $i=1${\tt c}, we have that
\begin{equation}
\langle f_{1{\tt c}} \rangle =
\sum_{j=1}^V P_j f_{1{\tt c}}(j) =
\sum_{\ell_2={\tt a}}^{{\tt z}} \sum_{\ell_3={\tt a}}^{{\tt z}}  \cdots 
P({\tt c}\ell_2\ell_3\dots ) 
= P^I_1({\tt c})
= \rho_1({\tt c})
\label{EqConstrain}
\end{equation}
(using that only words starting with $\ell_1={\tt c}$ contribute to the sum);
in words, we obtain that the expected value of the feature $1${\tt c}
is the marginal probability $P^I_1({\tt c})$ that the first letter in a word is {\tt c}, 
which we make equal to its empirical value $\rho_1({\tt c})$
(which is just the number of tokens with letter {\tt c} in position 1 divided by the total number of tokens).
Notice that we do not impose normalization constrain for the $P_j$'s,
as this is implicit in the marginals.
Coming back to the expression for the probabilities, 
Eq. (\ref{solution}), we have, for a three-letter example,
$$
P^{I}({\tt c a t})=\exp(\lambda_{1{\tt c}}+\lambda_{2{\tt a}}+\lambda_{3{\tt t}}-1),
$$
the label $I$ standing for 
the fact that the solution is obtained from the constrains of
one-letter marginals.
Substituting this into the constrain, Eq. (\ref{EqConstrain}),
we arrive to solutions of the form $e^{\lambda_{1{\tt c}}-1/3}=\rho_1({\tt c})$
and so,
$$
P^{I}({\tt c a t})
=\rho_1({\tt c}) \rho_2({\tt a}) \rho_3({\tt t})
$$
(note that other solutions for the $\lambda_{1{\tt c}}$'s are possible, 
but they lead to the same $P^I$'s;
in particular, the origin of each potential is not fixed
and one could replace, for instance, 
$\lambda_{1 \ell_1}\rightarrow 
  \lambda_{1 \ell_1}+C_1$ 
for all $\ell_1$, 
provided that the other potentials are modified accordingly
to yield the same value of the sum).
This model based on univariate (single-letter) marginals is very simple indeed, 
and closely related to monkey-typing models \cite{Miller_monkey,Ferrer-i-Cancho_2010}, 
as we obtain that each word is an independent combination of letters, 
with each letter having its own probability of occurrence
(but depending on its position in the word).

\subsection{Pairwise constrains}

The approach of Stephens and Bialek \cite{Stephens_Bialek} 
is the generalization of the previous model to two-letter features, 
which leads to constrains over the two-letter marginals. 
For instance, if the feature $i=12${\tt ca} denotes that the word
has letter {\tt c} in position 1 and letter {\tt a} in 2,
\begin{equation}
\langle f_{12{\tt ca}} \rangle =
\sum_{\forall j} P_j f_{12{\tt ca}}(j) =
\sum_{\ell_3={\tt a}}^{{\tt z}}  
\sum_{\ell_4={\tt a}}^{{\tt z}} 
\cdots 
P({\tt c a}\ell_3\dots)
=P^{II}_{12}({\tt c a})
=\rho_{12}({\tt c a}),
\label{tocite}
\end{equation}
with $\rho_{12}({\tt c a})$ the two-letter marginal, provided by the empirical data,
$$
\rho_{12}({\tt c a}) = \frac{\mbox{number of tokens with {\tt c} in 1 and {\tt a} in 2}}
{\mbox{total number of tokens}}.
$$
The solution (\ref{solution}),
restricted for the particular example of a three-letter word
can be written as
\begin{equation}
P^{II}({\tt c a t})=\exp(\lambda_{12}({\tt ca})+\lambda_{13}({\tt ct})+\lambda_{23}({\tt at})-1),
\label{Pcat}
\end{equation}
using the notation $\lambda_{12{\tt ca}}=\lambda_{12}({\tt ca})$ for
the multipliers, and
the label $II$ denoting that we are dealing with 
theoretical probabilities arising from
two-letter features, 
i.e., two-letter marginals. 
The same result writes, in general,
\begin{equation}
P^{II}({ \ell_1\ell_2\dots \ell_K})=
\exp\left(-1 + \sum_{k=1}^{K-1} \sum_{k'=k+1}^K\lambda_{k k'}(\ell_k \ell_{k'})\right),
\label{duda}
\end{equation}
with $K$ the word length (in number of letters).
Comparing with Boltzmann distribution, Eq.~(\ref{EqBoltzmann}), we can 
identify the Lagrange multiplier for each feature with the pairwise interaction potential
between the letters defining the feature 
(with a minus sign, and with a shift of one unit); 
for example,
$$
-\beta E({\tt c a t})=\lambda_{12}({\tt ca})+\lambda_{13}({\tt ct})+\lambda_{23}({\tt at})-1,
$$
and in general,
$$
-\beta E({ \ell_1\ell_2\dots \ell_K})= -1+
\sum_{k=1}^{K-1} \sum_{k'=k+1}^K\lambda_{k k'}(\ell_k \ell_{k'}).
$$
Therefore, words can be seen as {networks of interacting letters}
(with all-to-all interaction between pairs,
and where the position of the two letters matters for the interaction).
Note that three-letter interacions, common in English ortographic rules,
are not captured by the pairwise interaction;
for example, in positions $3$ to $5$: {\tt believe} (rule) versus {\tt deceive}
(exception, due to the {\tt c} letter).
Remarkably, this pairwise approach
has been used also for
{neuronal, biochemical, and genetic} networks \cite{Stephens_Bialek}.
A very simplified case of this letter system
turns out to be equivalent to an Ising model
(or, more properly, a spin-glass model):
just consider an alphabet of two letters ({\tt a} and {\tt b})
and impose the symmetries (not present in linguistic data, in general)
$\lambda_{k k'}({\tt a b})=\lambda_{k k'}({\tt b a})$
and 
$\lambda_{k k'}({\tt a a})=\lambda_{k k'}({\tt b b})$
(if one wants to get ride of this symmetries in the Ising system one could consider
external ``magnetic'' fields, associated to the one-letter marginals).

Substituting the solution (\ref{solution}) or (\ref{Pcat}) into the constrains (\ref{tocite}),
the equations we need to solve would be like
$$
P_{12}^{II}({\tt ca})=
\langle f_{12{\tt ca}}\rangle =
\sum_j f_{12{\tt ca}}(j) \,e^{\, -1 +\sum_{i=1}^m \lambda_i f_i(j)} =
$$
$$
=
e^{\lambda_{12}({\tt ca})} \sum_{\ell_3={\tt a}}^{\tt z}
e^{
\lambda_{13}({\tt c\ell_3})
+\lambda_{23}({\tt a\ell_3})
-1} = \rho_{12}({\tt ca}),
$$
if we restricted to three-letter words.

For computational limitations, we will only treat words comprising from $1$ to $6$ letters.
As the numerical algorithm we will use requires that the number of letters is constant
(see the Appendix), we will consider that words shorter than length $6$
are six-letter words whose last positions are filled with blanks; 
for example, {\tt cat $=$ cat$\square\square\square$}, 
where the symbol $\square$ denotes a blank.
In this way, instead of the usual $26$ letters in English we deal with $27$
(the last term in the sums of some of the previous equations 
should be $\square$, instead of {\tt z}).
This yields $6 \times 5/2=15$ interaction potentials ($15$ features) for each word, 
and a total of $15\times 27^2=10,935$ unknown values of the interaction potential
(i.e., Lagrange multipliers with minus sign)
corresponding to $10,935$ equations (one for each value of the two-letter marginals).
In contrast, note that there are about
$27^6=387,420,489$ 
possible words of lenght between $1$ and $6$ 
(the figures turn out to be a bit smaller
if one recalls that blanks can only be at the end of the word, 
in fact, $26+\dots + 26^6=321,272,406$).
In more generality, the $10,935$ equations to solve are like
\begin{equation}
e^{\lambda_{12}({\tt ca})} \sum_{\ell_3 \dots \ell_6}
e^{
\lambda_{13}({\tt c\ell_3})+\dots
+\lambda_{16}({\tt c\ell_6})
+\lambda_{23}({\tt a\ell_3})+\dots
+\lambda_{26}({\tt a \ell_6})
+\lambda_{34}({\tt \ell_3\ell_4})+\dots \dots
+\lambda_{56}({\tt \ell_5\ell_6})
-1} = \rho_{12}({\tt ca}),
\label{eqtosolve}
\end{equation}
where the solution is not straightforward anymore, 
and has to be found numericaly.
So, we deal with a constrained optimization problem, 
for which the Appendix provides complete information.
Here we just mention that the improved iterative-scaling method
consist in the successive application of transformations as
$$
\lambda_{12}({\tt ca}) \rightarrow \lambda_{12}({\tt ca})
+ \frac 1 6 \ln \frac{\rho_{12}({\tt ca})}{P_{12}^{II}({\tt ca})},
$$
see Eq. (\ref{EqAbove}) in the Appendix.
Note that, as in the case of univariate marginals, 
the potentials are undetermined under a shift, 
i.e., 
$\lambda_{12}(\ell_1 \ell_2)\rightarrow 
  \lambda_{12}(\ell_1 \ell_2)+C_{12}$, 
as long as the other potentials are correspondingly shifted
to give the same value for the sum.

\section{Data and results}

\subsection{Data}

As a corpus, we use all English books in 
the recently presented Standardized Project Gutenberg Corpus
\cite{Gerlach_Font_Clos}.
This comprises  
more than 40,000 books in English, 
with a total number of tokens
2,016,391,406
and a vocabulary size 
$V=2,268,043$.
The entropy of the corresponding word-probability distribution is $S=10.27$ bits.
In order to avoid spurious words (misspellings, etc.),
and also, 
for computational limitations, 
we disregard word types with absolute frequency smaller than 
10,000.
Also, word types (unigrams) 
containing characters different than the plain 26 letters from {\tt a} to {\tt z}
are disregarded
(note that we do not distinguish between capital and lower-case letters).
Finally, we remove also Roman numerals
(these are not words for our purposes, as
they are not formed by interaction between letters).
This reduces the number of tokens to
1,881,679,476 and $V$ to 11,042, and
so the entropy becomes $S=9.45$ bits.
Finally, the subset of words with length smaller or equal to 6
yields
1,597,358,419 tokens, $V=5,081$
and $S=8.35$ bits.
We will see that these sub-corpora fulfill Zipf's law, 
but each one with a slightly different power-law exponent.

%
%
%

\subsection{Marginal distributions}

\begin{figure}[ht]
\includegraphics[width=.40\columnwidth]{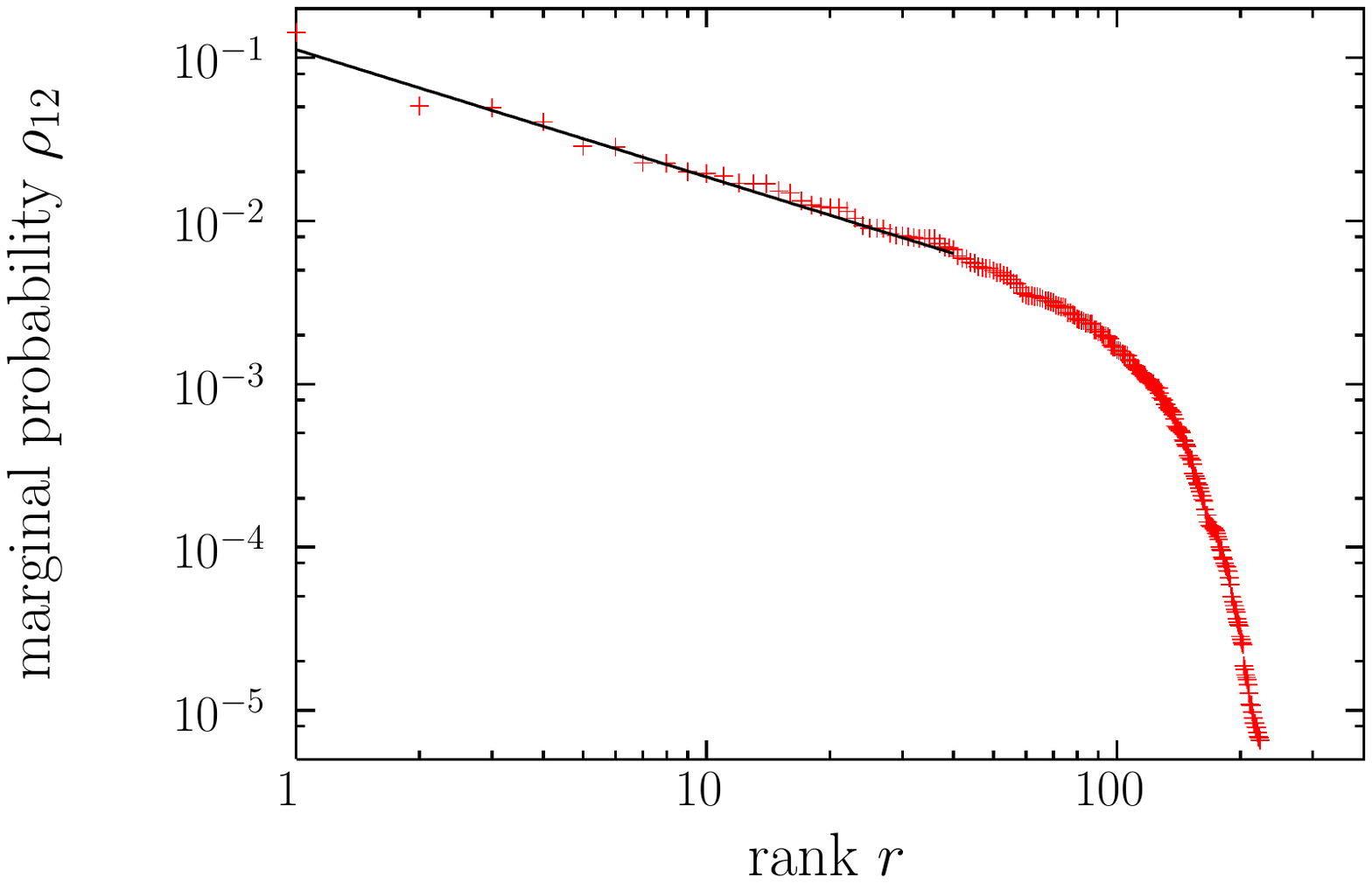}
\includegraphics[width=.40\columnwidth]{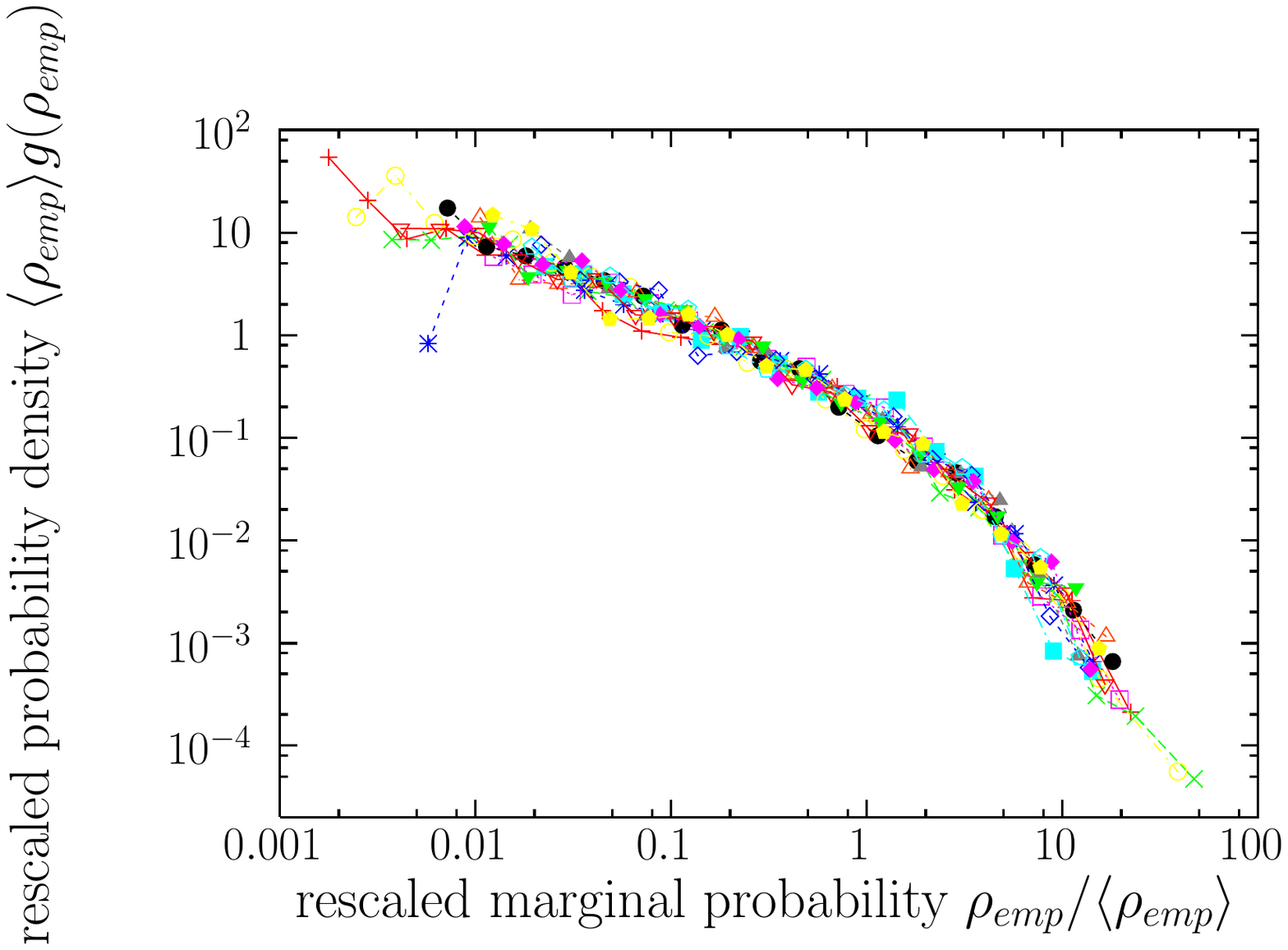}
\caption{
Empirical two-letter marginal distributions
(for word length not larger than 6 letters).
Left:
The distribution $\rho_{12}$ is 
represented in terms of the rank-frequency plot 
[corresponding to Eq. (\ref{rankfrequency})].
The most common values of $\rho_{12}$
correspond to the following pairs:
   {\tt th, an, of, to, he, in, a$\square$, ha, wh, wa, ...}
Power-law fit from Table \ref{table1} is shown as a straight line.
Right:
All 15 two-letter marginals
are represented in terms of the distributions of the value of the marginal probabilities,
$\rho_{12}, \rho_{13}, \dots \rho_{56}$ (denoted in general as $\rho_{emp}$).
All the distributions have been shifted (in log-scale) by rescaling by their mean values $\langle \rho _{emp}\rangle$,
see Ref. \cite{Corral_csf}. 
This makes apparent the similarities between all the two-letter marginal distributions, except for a scale factor given by $\langle \rho_{emp} \rangle$. 
Values below the mean ($\rho _{emp} <\langle \rho _{emp}\rangle$)
can be fitted by a truncated power law, with exponent $1+\beta \simeq 0.9$
(not reported in the tables).
}
\label{Fmarginals}
\end{figure}

Figure \ref{Fmarginals} displays the empirical two-letter marginal probabilities
(obtained from the 6-or-less-letter sub-corpus just described), 
which constitute the target of the optimization procedure.
There are a total of 5,092 non-zero values of the marginals.
Notice that, although the two-letter marginals are bivariate probabilities \cite{Stephens_Bialek},
Zipf's representation allows one to display them as univariated.
This is achieved by defining a rank variable, 
assigning rank $r=1$ to the type with the highest empirical frequency $\rho$
(i.e., the most common type),
$r=2$ to the second most common type, and so on (Fig. \ref{Fmarginals}(left)).
This is called the rank-frequency representation (or, sometimes, distribution of ranks).
Then, Zipf's law can be formulated as a power-law relation between $\rho$ and $r$,
\begin{equation}
\rho \propto \frac 1 {r^{1/\beta}}
\label{rankfrequency}
\end{equation}
for some range of ranks (typpically the lowest ones, i.e., the highest frequencies),
with the exponent $\beta^{-1}$ taking values close to one
(the symbol $\propto$ denotes proportionality).
When we calculate and report entropies we use always the rank-frequency representation. 

An approximated alternative representation (also used by Zipf) considers the empirical frequency $\rho$
as a random variable, whose distribution is computed.
In terms of the complementary cumulative distribution, $G(\rho)$, Zipf's law can be written as
\begin{equation}
G(\rho)  \propto \frac 1 {\rho^{\beta}},
\label{cumulative}
\end{equation}
which in terms of the probability density or probability mass function of $\rho$ leads to
\begin{equation}
g(\rho)  \propto \frac 1 {\rho^{\beta+1}},
\label{density}
\end{equation}
asymptotically, for large $\rho$
(Fig. \ref{Fmarginals}(right)).
Both $G(\rho)$ and $g(\rho)$ constitute a representation
in terms of the distribution of frequencies.
For more subttle arguments relating $\rho(r)$, $G(\rho)$, and $g(\rho)$, see Refs. \cite{Mandelbrot61,Moreno_Sanchez}.

We can test the applicability of Zipf's law to our two-letter marginals, 
in order to evaluate how surprising or unsurprising is the emergence from them of Zipf's law
in the word distribution.
Remember that, in the case of marginal distributions, types are pairs of letters.
Figure~\ref{Fmarginals} left shows that, despite the number of data in the marginals is relatively low 
(a few hundreds as shown in Table \ref{table1}, 
with a theoretical maximum equal to $26^2=676$),
the marginal frequencies appear as broadly distributed, 
varying along 4 orders of magnitude.
Although the double logarithmic plots do not correspond to straight lines,
the high-frequency (low-rank) part of each distribution can be fitted to a power law, 
for a number of orders of magnitude ranging from 0.5 to 2 
and an exponent $\beta$ typically between 1 and 2, as it can be seen in Table \ref{table1}.
Thus, the two-letter marginal distributions display a certain Zipfian character
(at least considering words of length not larger than 6, in letters),
with a short power-law range, in general,
and
with a somewhat large value of $\beta$
(remember that $\beta$ has to be close to one for the fulfilment of Zipf's law).

Remarkably, Fig. \ref{Fmarginals}(right) also shows
that all the marginal distributions present a characteristic, roughly the same 
shape, with the only difference being on the scale parameter of the frequency distribution,
which is determined by the mean frequency $\langle \rho_{k k'} \rangle$
(denoted generically in the figure as $\langle \rho_{emp} \rangle$).
This means, as shown in the figure, that
the distribution $g(\rho_{emp})$, when multiplied (rescaled) by 
$\langle \rho_{emp} \rangle$, can be considered, approximately, 
as a function that only depends of the rescaled frequency, 
$\rho_{emp} /  \langle \rho_{emp} \rangle$,
independently on which potential $\rho_{k k'}$ one is considering.
In terms on the distribution of ranks this scaling property translates 
into the fact that $\rho_{emp} /  \langle \rho_{emp} \rangle$
can be considered a function of only $r/V$.

For the fitting we have used the method proposed in Refs. \cite{Corral_Deluca,Corral_Gonzalez},
based on maximum-likelihood estimation and Kolmogorov-Smirnov goodness-of-fit testing.
This method lacks the problems presented in the popular Clauset et al.'s recipe
\cite{Clauset,Corral_nuclear,Voitalov_krioukov}.
The fitting method is applied to $\rho$ as a random variable 
(instead than to $r$ \cite{Altmann_Gerlach});
this choice presents a number of important advantages, as discussed in Ref. \cite{Corral_Cancho}. 
The outcome of the method is a estimated value of the exponent $\beta$
together with a value of $\rho$, denoted by $a$, from which the power-law fit,
Eqs. (\ref{cumulative}) and (\ref{density}), is non-rejectable 
(with a $p-$value larger than 0.20, by prescription). 
Although other distributions different than the power law
can be fitted to the marginal data (e.g., lognormal \cite{Corral_Gonzalez})
our purpose is not to find the best fitting distribution, 
but just to evaluate how much Zipf's power law
depends on a possible Zipf's behavior of the marginals.

\begin{widetext}
\begin{table}[ht]
\caption{
Results of power-law fitting of the form $g(\rho) \propto 1/\rho^{\beta+1}$ 
(for $a \le \rho\le b$)
applied to the 15 empirical two-letter marginal distributions (with $b=\infty$),
to the empirical word frequency $\rho_{word}$
and to the theoretical maximum-entropy solution $P^{II}$.
The empirical distribution for words of any length, $\rho_{all\,word}$,
is also shown, in order to compare it with $\rho_{word}$.
$V$ is the number of types (pairs of letters or words);
$\rho_{max}$ is the highest empirical frequency;
o.m. is the number of orders of magnitude in the fit, $\log_{10}(\rho_{max}/a)$;
$n$ is the number of types that enter into the power-law fit;
$\sigma$ is the standard error of the fitted exponent;
and $p$ is the $p-$value of the goodness-of-fit test.
The ratio $n/V$ ranges from 0.09 to 0.3.
Only words of length from 1 to 6 are taken into account.
Blanks are not considered in the marginals.
50 values of $a$ and $b$ (when $b$ is not fixed to $\infty$) 
are analyzed per order of magnitude, 
equally spaced in logarithmic scale.
$p-$values are computed from 1000 Monte Carlo simulations. 
Fits are considered non-rejectable if $p\ge 0.20$.
}
\begin{tabular}{ crc rcc rcc}
\hline
distribution    & $V$ & \phantom{0}$\rho_{max}$\phantom{0} & $a$ $(\times 10^{-4})$ 
& $b$ & o.m. & $n$ & $\beta \pm \sigma$ & $p$ \\
\hline
%
$\rho_{12}$ &       223 &   0.143 &   63.1 &   $\infty$ &   1.36 &        40 &        1.282$\pm$0.213  &  0.21\\
$\rho_{13}$  &       471 &   0.146 &   16.6 &  $\infty$ &      1.94 &       133 &        1.138$\pm$0.097  &  0.24\\
$\rho_{14}$  &       455 &   0.038 &   34.7 &   $\infty$ &     1.04 &        81 &        1.391$\pm$0.156  &  0.23\\
$\rho_{15}$  &       391 &   0.043 &   36.3 &   $\infty$ &     1.07 &        78 &        1.433$\pm$0.175  &  0.28\\
$\rho_{16}$  &       285 &   0.042 &   69.2 &   $\infty$ &      0.78 &        45 &        2.110$\pm$0.324  &  0.23\\
$\rho_{23}$ &       309 &   0.160 &   57.5 &   $\infty$ &   1.44 &        42 &        1.207$\pm$0.197  &  0.29\\
$\rho_{24}$  &       361 &   0.049 &   60.3 &   $\infty$ &    0.91 &        50 &        1.466$\pm$0.210  &  0.24\\
$\rho_{25}$  &       334 &   0.057 &   52.5 &   $\infty$ &    1.04 &        53 &        1.309$\pm$0.183  &  0.29\\
$\rho_{26}$  &       240 &   0.055 &   145.\phantom{5} &   $\infty$ &      0.58 &        21 &        2.576$\pm$0.627  &  0.22\\
$\rho_{34}$  &       330 &   0.048 &   83.2 &   $\infty$ &      0.76 &        36 &        1.764$\pm$0.340  &  0.41\\
$\rho_{35}$  &       371 &   0.039 &   50.1 &   $\infty$ &      0.89 &        57 &        1.359$\pm$0.190  &  0.28\\
$\rho_{36}$  &       273 &   0.045 &   75.9 &   $\infty$ &      0.78 &        44 &        1.935$\pm$0.298  &  0.32\\
$\rho_{45}$  &       278 &   0.051 &   87.1 &   $\infty$ &      0.77 &        35 &        1.579$\pm$0.270  &  0.33\\
$\rho_{46}$  &       244 &   0.044 &   100.\phantom{5} &   $\infty$ &      0.64 &        31 &        1.946$\pm$0.378  &  0.28\\
$\rho_{56}$  &       154 &   0.115 &   72.4 &  $\infty$ &    1.20 &        34 &        1.140$\pm$0.201  &  0.58\\
\hline

$\rho_{all\, word}$
&     11042 &   0.071 &   1.0 &   0.073 &  2.85 &       925 &        0.925$\pm$0.030  &  0.25\\%

$\rho_{word}$&      5081 &   0.084 &   0.5 &   0.087 &  
                                                                                                             3.20 &      1426 &        0.811$\pm$0.023  &  0.31\\

$P^{II}$ &      2174013 &   0.081 &   0.2 &   0.083 &  3.53 &      2947 &        0.886$\pm$0.017  &  0.38\\
                         
\hline
\end{tabular}
\label{table1} 
\end{table} 
\end{widetext}

\subsection{Word distributions}

\begin{figure}[ht]
\includegraphics[width=.80\columnwidth]{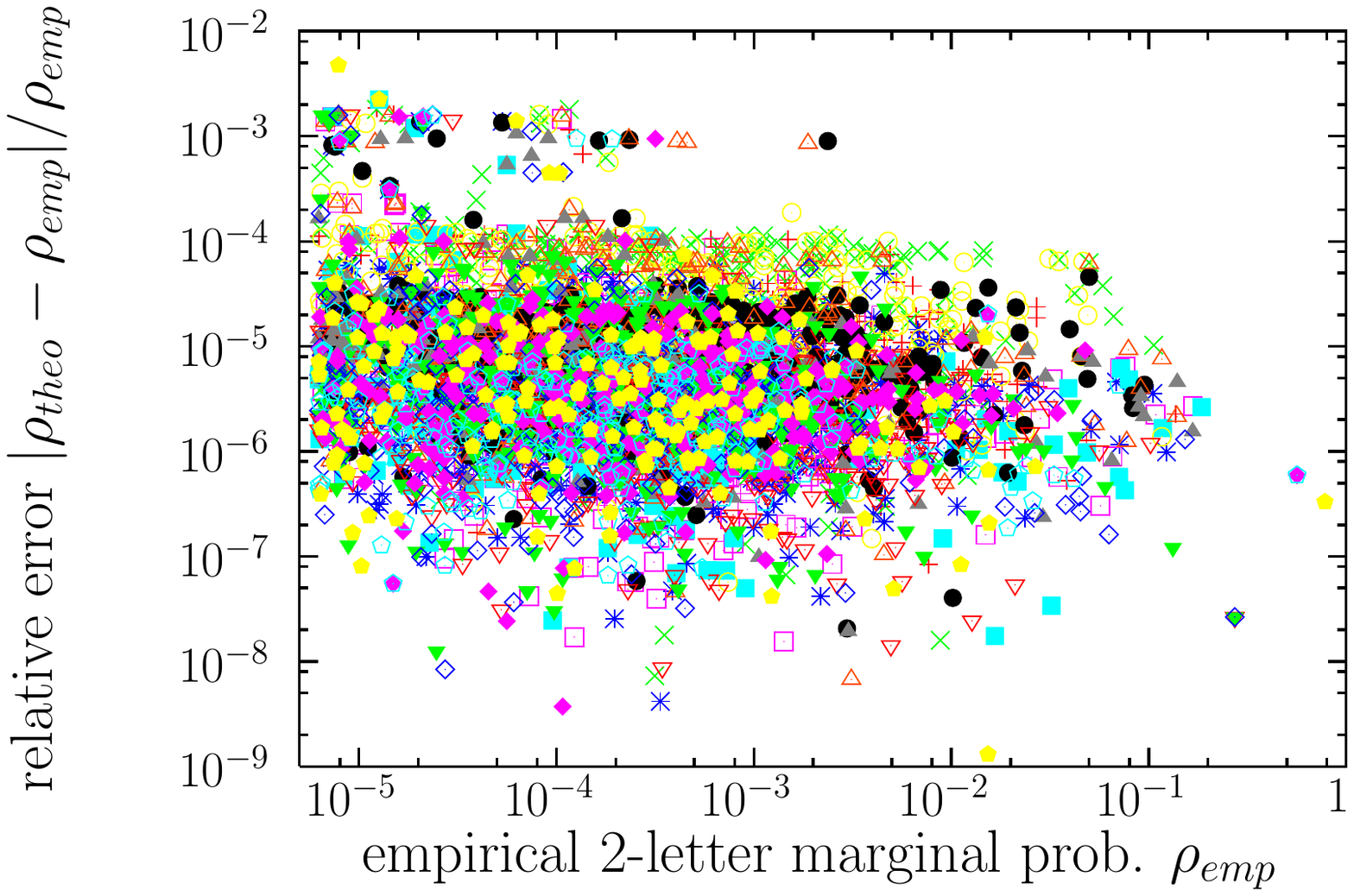}
\caption{
Comparison between the empirical two-letter marginal distributions $\rho_{emp}$
and the theoretical ones $\rho_{theo}$ obtained from the improved iterative-scaling optimization procedure \cite{Berger,Berger_etal}.
The relative error between both values of the marginal probability is shown as a function of the empirical value, for the 15 marginals.
}
\label{Ferror}
\end{figure}

\begin{figure}[ht]
\includegraphics[width=.40\columnwidth]{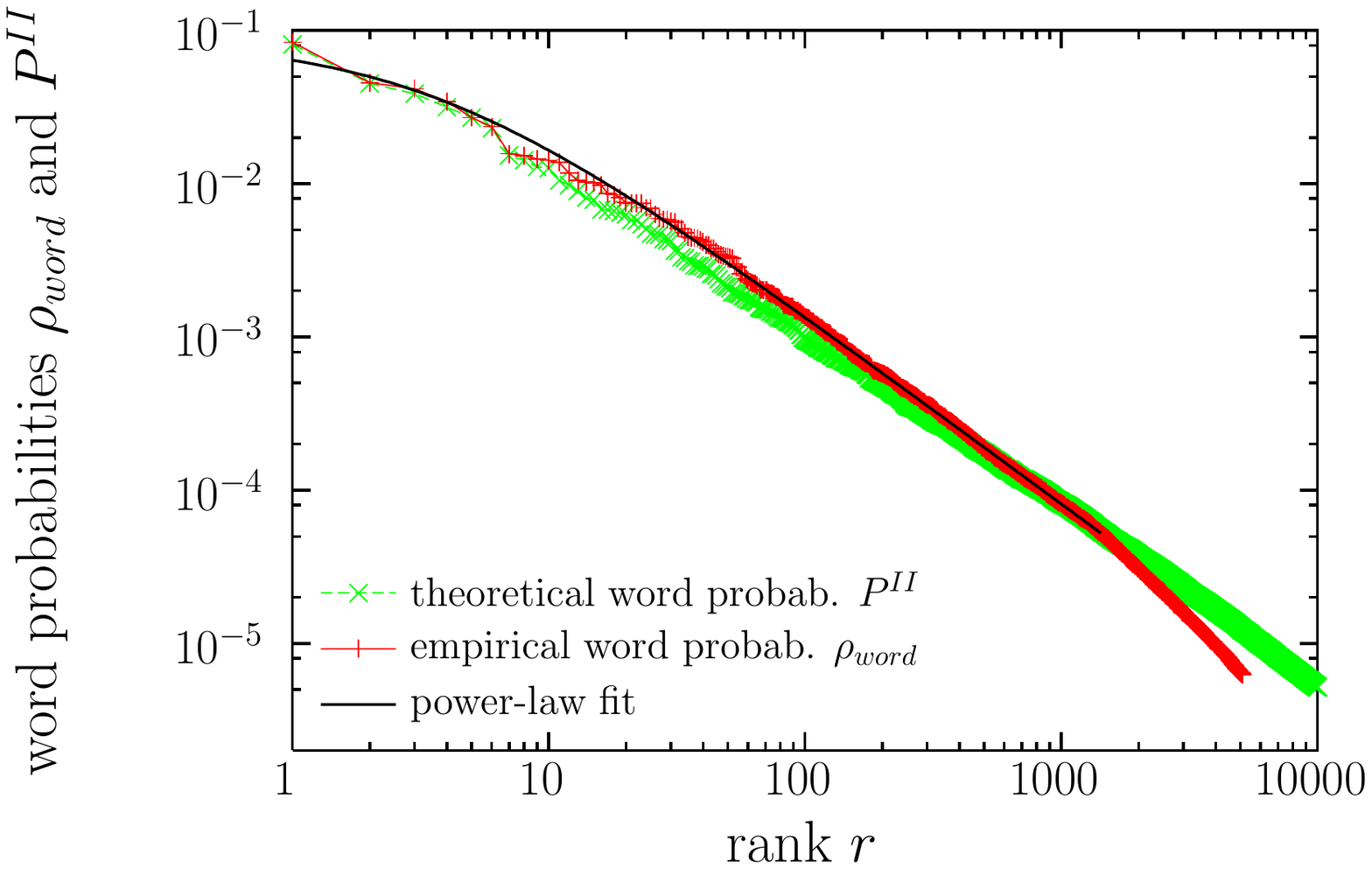}
\includegraphics[width=.40\columnwidth]{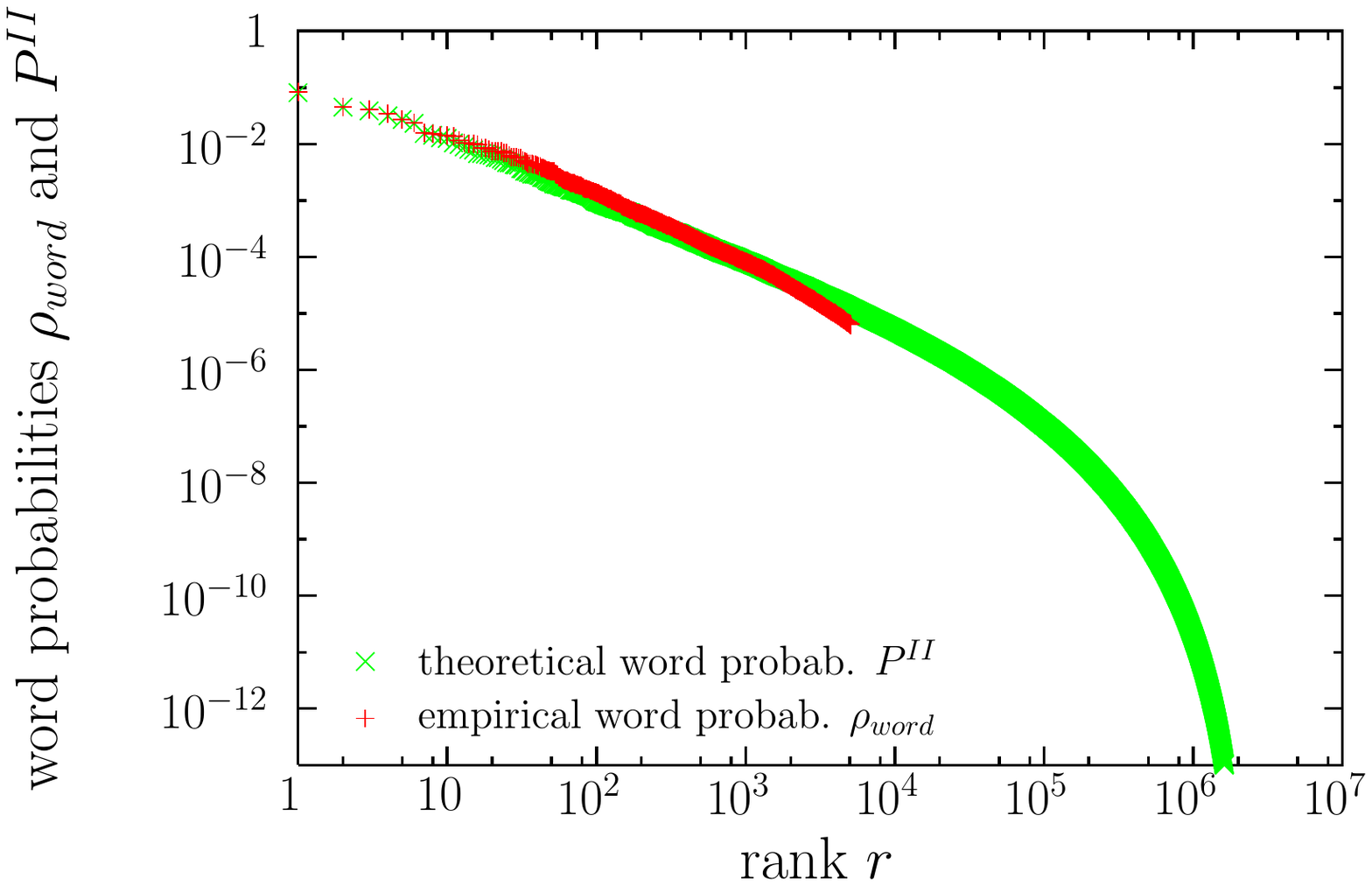}
\caption{
Empirical ($\rho_{word}$) and maximum-entropy theoretical ($P^{II}$)
word occurrence probabilities in the rank-frequency representation, 
together with the power-law fit of the distribution of frequencies
for the empirical case.
The same
distributions are shown at two different scales.
Left: only ranks below 10,000.
Right: only probabilities (frequencies) above $10^{-13}$.
}
\label{Fworddistribution}
\end{figure}

Figure \ref{Ferror} shows that the optimization succeeds in 
getting values of the theoretical marginal distributions 
very close to the empirical ones.
However, despite the fact the target of the optimization are the marginal distributions
(whose empirical values are the input of the procedure),
we are interested in the distribution of words, 
whose empirical value is known
but does not enter into the procedure, 
as this is the quantity we seek to ``explain''.
Zipf's rank-frequency representation allows us to display in one dimension
the six-dimensional nature
(from our point of view) of the word frequencies;
for the empirical word frequencies this is shown in Fig. \ref{Fworddistribution}.
We find that the distribution is better fitted in terms of an upper truncated power law \cite{Burroughs_Tebbens,Corral_Deluca}, 
given, as in Eq. (\ref{density}), 
by $g(\rho)  \propto  1/ {\rho^{\beta+1}}$ but in a finite range $a \le \rho \le b < \infty$ (the untruncated case would be recovered by taking $b\rightarrow \infty$).
This corresponds, in the continuum case, to a cumulative distribution 
$G(\rho)\propto 1/\rho^\beta - 1/b^\beta$, 
and to a rank-frequency relation
$$
\rho \propto \frac 1 {(r+V/b^\beta)^{1/\beta}},
$$
which coincides in its mathematical expression with the so-called Zipf-Mandelbrot distribution 
(although the continuous fit makes $r$ a continuous variable;
remember that $V$ is the number of types).
The fitting procedure is essentially the same as the one for the
untruncated power law outlined in the previous subsection,
with the maximization of the likelihood a bit more involved \cite{Corral_Deluca,Corral_Gonzalez}.

In Fig. \ref{Fworddistribution} we also display 
the theoretical result, $P^{II}$, Eq. (\ref{duda}),
arising from the solution of Eq. (\ref{eqtosolve}).
We see that, qualitatively, $P^{II}$
has a shape rather similar to the empirical one.
Both distributions fulfill Zipf's law, with exponents
$\beta$ equal to 0.89 and 0.81, respectively.
We also see in the figure that the quantitative agreement in the values of the probability ($P^{II}$ and $\rho_{word}$) is rather good for the smallest values of the 
rank ($r<10$); 
however, both curves start to slightly depart from each other for $r>10$.
In addition, the rank values are associated to the same word types 
for $r\le 6$ ({\tt the, of, and, to, a, in}), but for larger ranks the correspondence 
may be different ($r=7$ corresponds to {\tt i} in one case and to {\tt that} in the other).
If we could represent $\rho_{word}$ and $P^{II}$ in six dimensions
(instead that as a function of the rank)
we would see more clearly the differences between both.

\begin{figure}[ht]
\includegraphics[width=.80\columnwidth]{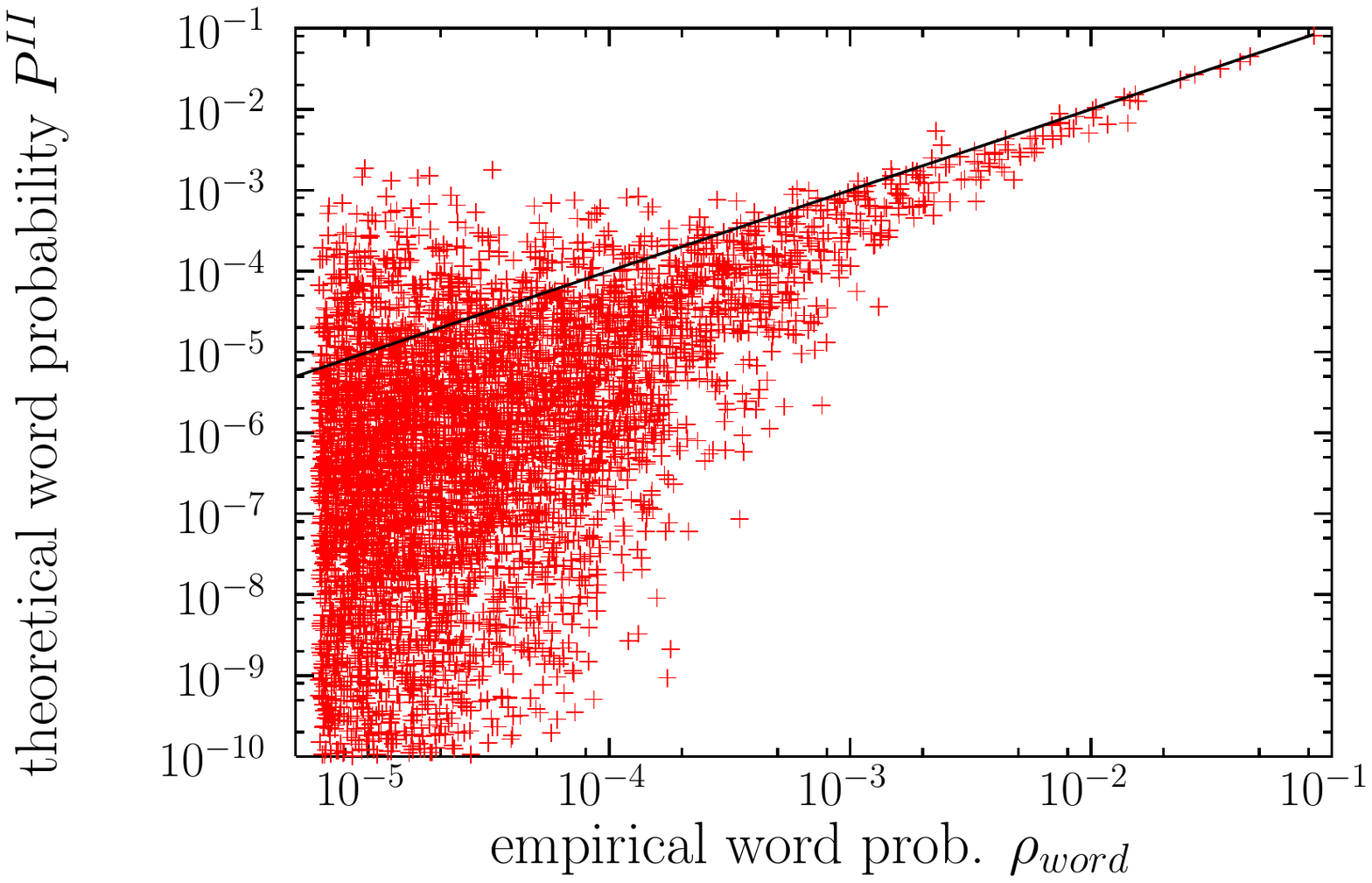}
\caption{
Maximum-entropy theoretical probability $P^{II}$ for each word type in the sub-corpus
as a function of its empirical probability (relative frequency) $\rho_{word}$.
The straight line would signal a perfect correpondence between 
$P^{II}$ and $\rho_{word}$.
Values of $P^{II}$ below $10^{-10}$ are not shown.
Words with the lowest $P^{II}$ (in the range $10^{-17}$--$10^{-15}$)
are {\tt
shaggy,
isaiah,
leslie,
feudal,
caesar,
yankee,
opium, 
yields,
phoebe,
sydney.
}
}
\label{Fscatterplot}
\end{figure}

Zipf's law is, in part, 
the reason of this problem, as for $r\ge 10$ the difference 
in probabilities for consecutive ranks becomes smaller than 10 \%,
see Eq. (\ref{rankfrequency}),
and for $r\ge 100$ the difference decreases to less than 1 \%
(assuming $\beta\simeq 1$).
So, finite resolution in the calculation of $P^{II}$
will lead to the ``mixing of the ranks.''
However, the main part of the problem comes from the unability of the algorithm
in some cases
to yield values of $P^{II}$ close to the empirical value, $\rho_{word}$, 
as it can be seen in the scatter plot of 
Fig. \ref{Fscatterplot} (In agreement with Ref. \cite{Stephens_Bialek}).
The entropy of the theoretical word probabilities turns out to be
$S=9.90$ bits, somewhat larger than the corresponding empirical value $8.35$ bits.
If we truncate this distribution, eliminating probabilities below 
$10,000/1,597,358,419\simeq 6\times 10^{-6}$ (as in the empirical distribution)
we get $S=8.88$ bits, still larger than the empirical value.
Existing (empirical) words for which the algorithm yields the lowest
theoretical probabilities are enumerated in the caption of the figure.
Curiously, as it can be seen, these are not particularly strange words.

An interesting issue is that the maximum-entropy solution, Eq. (\ref{duda}),
leads to the ``discovery'' of new words.
Indeed, whereas the empirical corpus has $V=5,081$ (number of types), 
the theoretical solution leads to $V=2,174,013$.
Most of these new words have very small probabilities;
however, there are others far from being rare (theoretically).
In this way, the most common theoretical word not present in the empirical corpus
is {\tt whe}, with a theoretical rank $r=40$
(it should be the $40-$th most common word in English, 
for length six or below,
following the maximum-entropy criterion).
Table \ref{table2} provides the first 25 of these new words, 
ranked by their theoretical probability $P^{II}$.
We see that the ortography of these words looks very ``reasonable''
(they look like true English words).
On the other side, the most rare words, with probability 
$P^{II} \sim 10^{-30}$, are nearly impossible English words, as:
{\tt sntnut, ouoeil, oeoeil, sntnu, snsnua}... (not in the table).

\begin{widetext}
\begin{table}[ht]
\caption{
Most common
theoretical words from the maximum-entropy procedure
that are not present in the analyzed sub-corpus.
In fact, some of these words are present in the original complete corpus
(may be as misspellings),
but not in our sub-corpus (as we have disregarded frequencies smaller than 10,000).
$r$ is (theoretical) rank and
$P^{II}$ is (theoretical) probability.
}
\begin{tabular}{ ccl }
\hline
$r $ & $P^{II}$ & word \\
\hline
          40 &   2.88$\times 10^{-3}$ & whe   \\
          48 &   2.20$\times 10^{-3}$ & wis   \\
          52 &   1.95$\times 10^{-3}$ & mo    \\
          61 &   1.74$\times 10^{-3}$ & wast  \\
          64 &   1.69$\times 10^{-3}$ & ond   \\
          71 &   1.52$\times 10^{-3}$ & ar    \\
          77 &   1.40$\times 10^{-3}$ & ane   \\
          87 &   1.24$\times 10^{-3}$ & ald   \\
          89 &   1.21$\times 10^{-3}$ & bo    \\
          92 &   1.16$\times 10^{-3}$ & thes  \\
          94 &   1.10$\times 10^{-3}$ & hime  \\
          98 &   9.83$\times 10^{-4}$ & hive  \\
         102 &   9.45$\times 10^{-4}$ & thise \\
         103 &   9.39$\times 10^{-4}$ & af    \\
         110 &   8.80$\times 10^{-4}$ & wer   \\
         117 &   8.16$\times 10^{-4}$ & thay  \\
         118 &   8.16$\times 10^{-4}$ & hes   \\
         123 &   7.88$\times 10^{-4}$ & wath  \\
         125 &   7.82$\times 10^{-4}$ & hor   \\
         127 &   7.60$\times 10^{-4}$ & sime  \\
         134 &   7.22$\times 10^{-4}$ & tome  \\
         135 &   7.21$\times 10^{-4}$ & har   \\
         141 &   6.94$\times 10^{-4}$ & thit  \\
         143 &   6.86$\times 10^{-4}$ & mas   \\
         146 &   6.77$\times 10^{-4}$ & hew   \\                     
\hline
\end{tabular}
\label{table2} 
\end{table} 
\end{widetext}

\subsection{Values of Lagrange multipliers and potentials 
}

\begin{figure}[ht]
\includegraphics[width=.40\columnwidth]{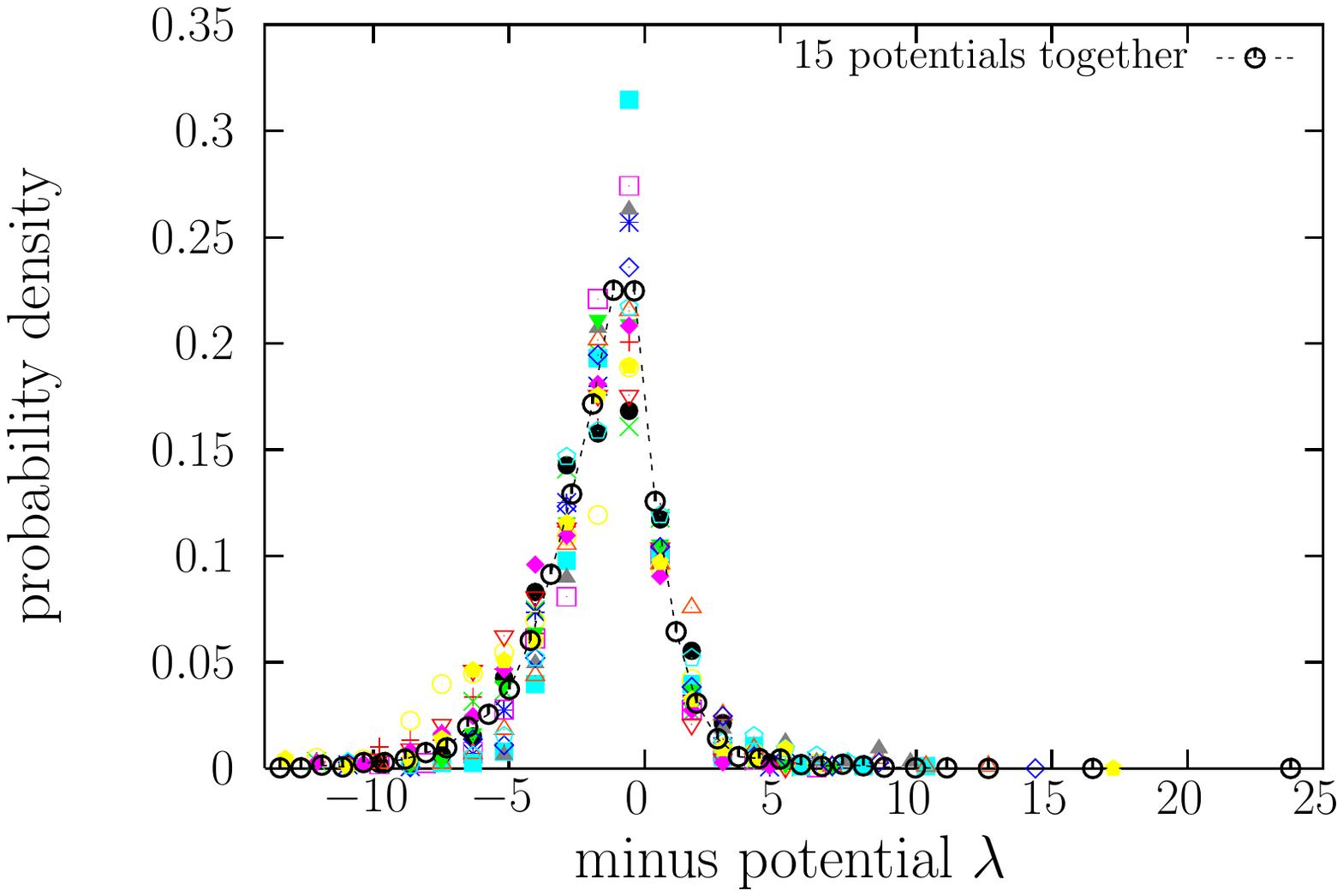}
\includegraphics[width=.40\columnwidth]{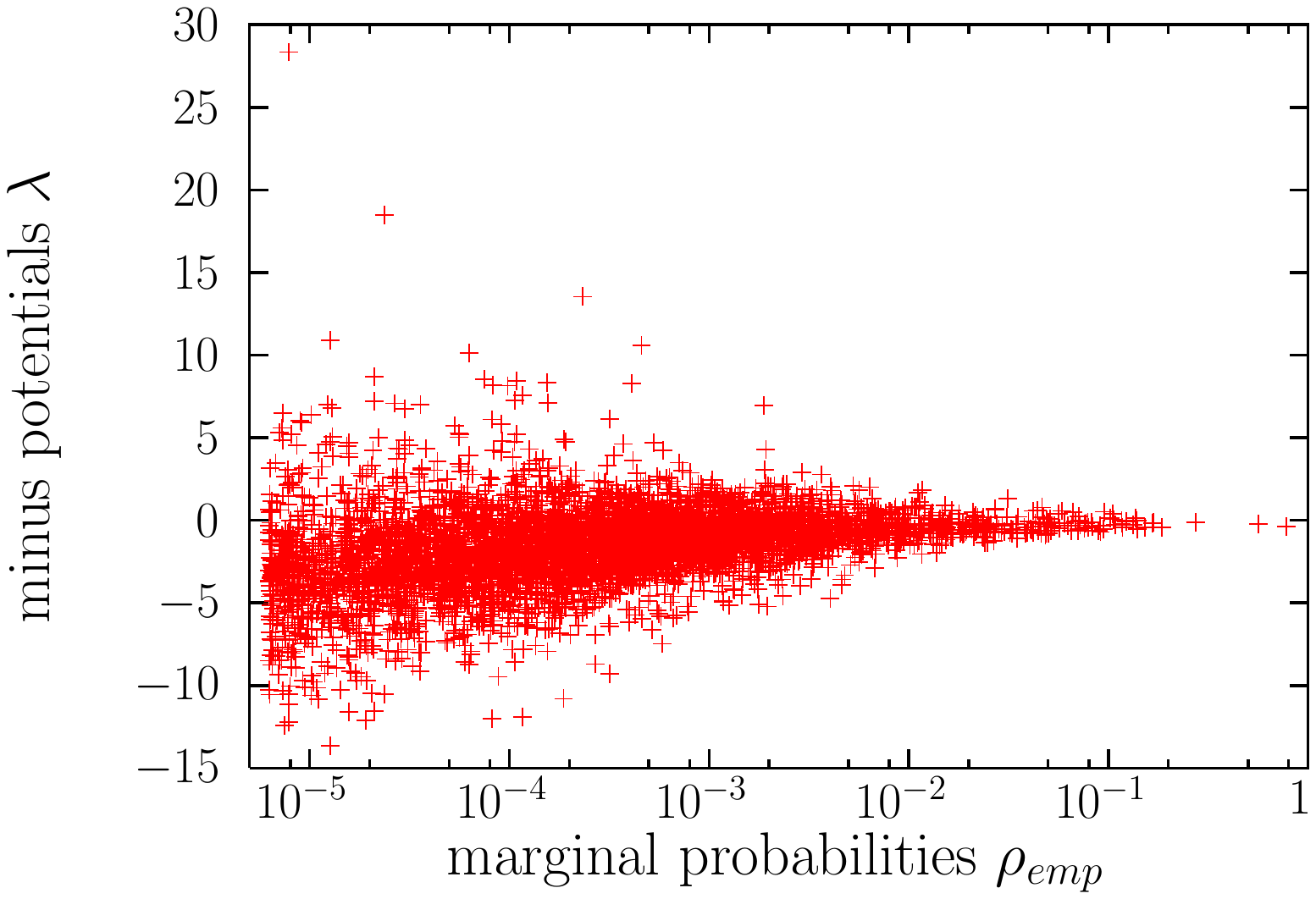}
\caption{
Left: 
Empirical probability densities of the 15 individual potentials (with a negative sign)
and the probability density of the 15 agregated data sets.
Right: 
Value of the Lagrange multiplier
(which corresponds to the interaction potential with a negative sign)
for each pair of letters (and positions)
as a function of the corresponding marginal probability.
}
\label{Fpotentials}
\end{figure}

We have established that, for a given word, the value of its occurrence probability $P^{II}$
comes from the exponentiation of the sum the 15 interaction potentials between the
6 letter positions that constitute the word (in our maximum-entropy approach).
So, the values of the potentials (or the values of the Lagrange multipliers) determine the value 
of the probability $P^{II}$.
It is interesting to investigate, given a potential or a multiplier 
(for instance $\lambda_{12}$),
how the different values it takes ($\lambda_{12}({\tt a a}), \lambda_{12}({\tt a b})$, etc.)
are distributed.
Curiously, we find that the 15 different potentials are (more or less) equally distributed, 
i.e., follow the same skewed and spiky distribution, as shown in Fig. \ref{Fpotentials}(left).

One can try to use this fact to shed some light on the origin of Zipf's law.
Indeed, exponentiation is a mechanism of power-law generation
\cite{Sornette_critical_book,Corral_nuclear}.
We may arguee that the sum of 15 random numbers drawn from the same
spiky distribution has to approach, by the central limit theorem, a normal distribution, 
and therefore, the exponentiation of the sum would yield a lognormal distribution for $P^{II}$
(i.e., a lognormal shape for $g(P^{II})$).
However, this may be true for the central part of the distribution, but
not for its rightmost extreme values, which is the part of the distribution we are more interested in
(high values of $P^{II}$, i.e., the most common words).
Note also that, in practice, 
for calculating the probability of a word, we are not summing 15 equally distributed independent random numbers, as not all the words are possible;
i.e., there are potentials that take a value equal to infinite, due to forbidden combinations,
and these infinite values are not taken into account in the distribution of the potentials.
An additional problem with this approach is that, although most values of the potentials converge to a 
fix value (and the distribution of potentials shown in the figure is stable), there are single values
that do not converge, related to words with very low probability. 
These issues need to be further investigated in future research.
In addition,
Fig. \ref{Fpotentials}(right) shows, as a scatter plot, 
the dependence between the value of each potential and the corresponding two-letter marginal probability.
Although Eq. (\ref{eqtosolve}) seems to indicate a rough proportionality
between both, the figure shows that such proportionality does not hold
(naturally, the rest of terms in the equation play their role).

\section{Discussion}

We have generalized a previous study of Stephens and Bialek \cite{Stephens_Bialek}.
Instead of restricting our study to four-letter words, 
we consider words of any length from one to six, 
which leads to greater computational difficulties,
and employ a much larger English corpus as well.
We perform an analysis of the fulfilment of Zipf's law
using state-of-art statistical tools.
Our more general results are nevertheless 
in the line of those of Ref. \cite{Stephens_Bialek}.
We see how the frequency of occurrence of pairs of letters in words
(the pairwise marginal distributions), 
together with the maximum-entropy principle
(which provides the distribution with the maximum possible randomness),
constrain the probabilities of word occurrences in English. 

Regarding the shape of the distributions,
the agreement between the maximum-entropy solution for the word distribution
and its empirical counterpart is very good at the qualitative level,
and reasonably good at the quantitative level for the most common words,
as shown in Fig. \ref{Fworddistribution}.
Moreover, new possible English words, not present in the corpus 
(or, more exactly, in the subcorpus we have extracted) have been ``discovered'',
with hypothetical (theoretical) values of the occurrence probability
that vary along many orders of magnitude.
However, regarding the probabilities of concrete words, 
the method yields considerable scatter of the theoretical probabilities
(in comparison with the known empirical probabilities), 
except for the most common words, see Fig. \ref{Fscatterplot}.

As two by-products, we have found that the pairwise (two-letter) occurrence 
distributions are all characterized by a well defined shape, 
see Fig. \ref{Fmarginals}(right), 
and that the distributions of the 15 different interaction potentials
are nearly the same, see Fig. \ref{Fpotentials}(left).
The latter is an intringuing fact that we have tried to relate, without success yet,
to other skewed and spiky distributions that appear in complex and correlated systems,
such as the so-called 
Bramwell-Holdsworth-Pinton (BHP) distribution \cite{BHP},
the Tracy-Widom distribution, or
the Kolmogorov-Smirnov distribution \cite{Font_clos_molon,Corral_garcia_moloney_font}.

All-to-all interaction between six elements (six letter positions) leads to 15 pairs, 
so to 15 interaction potentials. 
One may argue if the all-to-all interaction is realistic 
(for example, up to which point the first letter influences the last one).
The fact that the values that the interaction potentials take
are more or less the same for all of them (Fig. \ref{Fpotentials}(left))
indicates that all potentials are equally important.
Nevertheless, one could abandon the all-to-all interaction
and embrace instead nearest-neighbor coupling.
This reduces the number of potentials from 15 to 5
(with open boundary conditions),
with the subsequent computational simplification.
A further reduction would be to impose that all potentials
are the same (i.e., they do not depend on letter positions, only on diffence of positions, 
e.g., $\lambda_{12}=\lambda_{23}$, etc.).
This leads to only one potential (in the case of nearest-neighbor interaction;
5 potentials in the all-to-all case).
It would be interesting to see how these modifications compare with the original model;
this is left for future research.
An extension towards a different direction would be to use 
phonemes or syllables instead of letters as the constituents of words.
We urge the authors of the corpus in Ref. \cite{Gerlach_Font_Clos}
to provide the decomposition of the words in the corpus into these parts.
Remarkably, the approach presented here, and in Ref. \cite{Stephens_Bialek}
has also been applied to music 
\cite{Tria_Loreto_music}.

\section{Acknowledgements}

We are grateful to J. Davidsen for revealing to us the existence of
Ref. \cite{Stephens_Bialek} and to
F. Font-Clos and M. Gerlach for providing, prior to publication, 
the database created in Ref. \cite{Gerlach_Font_Clos}.
Irina Espejo participated in the early stages of this project;
we are indebted to her specially for drawing our attention 
to Refs. \cite{Berger_etal,Berger}.
Some preliminary results of this research were presented at the workshop
``Statistics of Languages'' (Warsaw, July 2017);
we are grateful to the organizers.
This work was largely completed at l'Abadia de Burg, Pallars Sobir\`a, Lleida.

\section{Appendix}

We summarize here the main formulas in Ref. \cite{Berger},
for the improved iterative-scaling method.
The per-datum {log-likelihood} $L(\vec\lambda)$ of the model $P_j(\vec\lambda)$ 
(stressing the dependence on the value of the set of parameters $\vec \lambda$) is
given by
$$
L(\vec\lambda)= \sum_{j=1}^V \rho(j) \ln P_j(\vec\lambda),
$$
with $\vec\lambda = (\lambda_1,\lambda_2,\dots \lambda_m)$ 
and $\rho(j)$ the empirical probability for word type $j$.
Substituting the maximum-entropy solution for the theoretical probability
Eq. (\ref{solution}),
written as 
$P_j = e^{\sum_i \lambda_i f_i(j)}/Z$ with $Z=\sum_j e^{\sum_i \lambda_i f_i(j)}$, 
one gets
$$
L(\vec\lambda)= \sum_{i=1}^m \lambda_i F_i -\ln Z,
$$
with $\sum_j \rho(j) f_i(j) = F_i$,
from Eq. (\ref{EqConstrain0}),
which leads to
$$
\frac{\partial L}{\partial \lambda_i}
=F_i -\frac 1 Z \frac 
{\partial }{\partial \lambda_i}\sum_j e^{\sum_{i'} \lambda_{i'} f_{i'}(j)}
= F_i -\langle f_i \rangle,
$$
using Eq. (\ref{EqConstrain0}).
This indicates that the parameters $\vec \lambda$ that {fulfill the constrains} also
{maximize the log-likelihood}, and vice versa, and therefore
the maximum-entropy parameters can be obtained
from maximum likelihood.

It can be shown that, for a change $\vec \delta$
in the values of the parameters, the increase in 
log-likelihood fulfils
$$
L(\vec\lambda+\vec\delta) - L(\vec\lambda) \ge 
\sum_j \rho(j) \sum_{i=1}^m \delta_i f_i(j) +1-\sum_j P_j(\vec\lambda) \sum_{i=1}^m \frac{f_i(j)}{ n(j)} e^{\delta_i  n(j)}
$$
with $ n(j) =\sum_i f_i(j)=$ number of features of word $j$. 
Now one should 
look for the values of $\vec \delta$
that {maximize the lower bound}
(right-hand side of the previous inequality).
Curiously, 
Ref. \cite{Berger}
does not provide the final solution,  
but this is in
Ref. \cite{Berger_etal} instead.
Maximizing, one gets
\begin{equation}
\delta_i= \frac 1  n \ln \frac{\sum_j \rho(j) f_i(j)}{\sum_j P_j(\vec\lambda) f_i(j)}
= \frac 1  n \ln  \frac{F_i}{\langle f_i \rangle}
\label{EqAbove}
\end{equation}
using that $ n(j) = $ constant $=  n$, if {word length is constant}
($6$ in our case, considering that blanks complete shorter words).
The
improved {iterative-scaling} algorithm is just:
Initialize $\lambda_i$, 
calculate $P_j(\vec\lambda)$ [Eq. (\ref{solution})], 
update ${\langle f_i \rangle}$ [Eq. (\ref{EqConstrain0})],
calculate $\delta_i$  [Eq. (\ref{EqAbove})] 
and the new $\lambda_i$ as $\lambda_i +\delta_i$, and so on.

As the equation to solve, Eq. (\ref{eqtosolve}), is a sum of exponentials, 
when a marginal value is not present in the empirical data, 
i.e., 
when the right-hand side of Eq. (\ref{eqtosolve}) is zero,
the left-hand side of the equation cannot verify the equality 
unless some Lagrange multiplier is minus infinite, 
which is a value that the numerical algorithm 
cannot achieve.
We therefore take from the beginning
the corresponding multiplier to be equal to minus infinity
(i.e., interaction potential equal to infinite).
To be concrete, if for example $\rho_{12}({\tt z z})=0$, 
we take $\lambda_{12}({\tt z z})=-\infty$, 
which leads to $P_j=0$ for any $j=\{{\tt z z} \ell_3 \ell_4 \dots\}$.
This means that we can
restrict our analysis of possible words to those 
with all pairs of letters corresponding to non-null empirical marginals, because
the rest of words have zero probability.

%

\begin{thebibliography}{10}

\bibitem{Li02}
W.~Li.
\newblock Zipf's law everywhere.
\newblock {\em Glottom.}, 5:14--21, 2002.

\bibitem{Malevergne_Sornette_umpu}
Y.~Malevergne, V.~Pisarenko, and D.~Sornette.
\newblock Testing the {Pareto} against the lognormal distributions with the
  uniformly most powerful unbiased test applied to the distribution of cities.
\newblock {\em Phys. Rev. E}, 83:036111, 2011.

\bibitem{Clauset}
A.~Clauset, C.~R. Shalizi, and M.~E.~J. Newman.
\newblock Power-law distributions in empirical data.
\newblock {\em SIAM Rev.}, 51:661--703, 2009.

\bibitem{Axtell}
R.~L. Axtell.
\newblock Zipf distribution of {U.S.} firm sizes.
\newblock {\em Science}, 293:1818--1820, 2001.

\bibitem{Pueyo}
S.~Pueyo and R.~Jovani.
\newblock Comment on ``{A} keystone mutualism drives pattern in a power
  function''.
\newblock {\em Science}, 313:1739c--1740c, 2006.

\bibitem{Camacho_sole}
J.~Camacho and R.~V. Sol\'e.
\newblock Scaling in ecological size spectra.
\newblock {\em Europhys. Lett.}, 55:774--780, 2001.

\bibitem{Adamic_Huberman}
L.~A. Adamic and B.~A. Huberman.
\newblock {Zipf's} law and the {Internet}.
\newblock {\em Glottom.}, 3:143--150, 2002.

\bibitem{Furusawa2003}
C.~Furusawa and K.~Kaneko.
\newblock Zipf's law in gene expression.
\newblock {\em Phys. Rev. Lett.}, 90:088102, 2003.

\bibitem{Zanette_music}
D.~H. Zanette.
\newblock {Zipf}'s law and the creation of musical context.
\newblock {\em Mus. Sci.}, 10:3--18, 2004.

\bibitem{Haro}
M.~Haro, J.~Serr\`a, P.~Herrera, and A.~Corral.
\newblock Zipf's law in short-time timbral codings of speech, music, and
  environmental sound signals.
\newblock {\em PLoS ONE}, 7:e33993, 2012.

\bibitem{Serra_scirep}
J.~Serr\`a, A.~Corral, M.~Bogu{\~n}\'a, M.~Haro, and J.~Ll. Arcos.
\newblock Measuring the evolution of contemporary western popular music.
\newblock {\em Sci. Rep.}, 2:521, 2012.

\bibitem{Baayen}
H.~Baayen.
\newblock {\em Word Frequency Distributions}.
\newblock Kluwer, Dordrecht, 2001.

\bibitem{Baroni2009}
M.~Baroni.
\newblock Distributions in text.
\newblock In A.~L\"udeling and M.~Kyt\"o, editors, {\em Corpus linguistics: An
  international handbook, Volume 2}, pages 803--821. Mouton de Gruyter, Berlin,
  2009.

\bibitem{Zanette_book}
D.~Zanette.
\newblock Statistical patterns in written language.
\newblock {\em arXiv}, 1412.3336v1, 2014.

\bibitem{Piantadosi}
S.~T. Piantadosi.
\newblock Zipf's law in natural language: a critical review and future
  directions.
\newblock {\em Psychon. Bull. Rev.}, 21:1112--1130, 2014.

\bibitem{Altmann_Gerlach}
E.~G. Altmann and M.~Gerlach.
\newblock Statistical laws in linguistics.
\newblock In M.~D. Esposti, E.~G. Altmann, and F.~Pachet, editors, {\em
  Creativity and Universality in Language. {Lecture Notes in Morphogenesis}}.
  Springer, 2016.

\bibitem{Moreno_Sanchez}
I.~Moreno-S\'anchez, F.~Font-Clos, and A.~Corral.
\newblock Large-scale analysis of {Zipf}'s law in {English} texts.
\newblock {\em PLoS ONE}, 11(1):e0147073, 2016.

\bibitem{Mitz}
M.~Mitzenmacher.
\newblock A brief history of generative models for power law and lognormal
  distributions.
\newblock {\em Internet Math.}, 1 (2):226--251, 2004.

\bibitem{Newman_05}
M.~E.~J. Newman.
\newblock Power laws, {Pareto} distributions and {Zipf}'s law.
\newblock {\em Cont. Phys.}, 46:323 --351, 2005.

\bibitem{Loreto_urn}
V.~Loreto, V.~D.~P. Servedio, S.~H. Strogatz, and F.~Tria.
\newblock Dynamics on expanding spaces: Modeling the emergence of novelties.
\newblock In M.~Degli~Esposti et~al., editor, {\em Creativity and Universality
  in Language}, pages 59--83. Springer, Switzerland, 2016.

\bibitem{Miller_monkey}
G.~A. Miller.
\newblock Some effects of intermittent silence.
\newblock {\em Am. J. Psychol.}, 70(2):311--314, 1957.

\bibitem{Ferrer-i-Cancho_2010}
R.~Ferrer~i Cancho and B.~Elvev{\aa}g.
\newblock Random texts do not exhibit the real {Zipf}'s law-like rank
  distribution.
\newblock {\em PLoS ONE}, 5(3):e9411, 03 2010.

\bibitem{Ferrer2002a}
R.~{Ferrer i Cancho} and R.~V. Sol\'e.
\newblock Least effort and the origins of scaling in human language.
\newblock {\em Proc. Natl. Acad. Sci. U.S.A.}, 100:788--791, 2003.

\bibitem{Prokopenko}
M.~Prokopenko, N.~Ay, O.~Obst, and D.~Polani.
\newblock Phase transitions in least-effort communications.
\newblock {\em J. Stat. Mech.}, 2010(11):P11025, 2010.

\bibitem{Dickman_Moloney_Altmann}
R.~Dickman, N.~R. Moloney, and E.~G. Altmann.
\newblock {Analysis of an information-theoretic model for communication}.
\newblock {\em J. Stat. Mech: Theory Exp.}, P12022, 2012.

\bibitem{Corominas_dice}
B.~Corominas-Murtra, R.~Hanel, and S.~Thurner.
\newblock Understanding scaling through history-dependent processes with
  collapsing sample space.
\newblock {\em Proc. Natl. Acad. Sci. USA}, 112(17):5348--5353, 2015.

\bibitem{Ferrer_cancho_compression}
R.~{Ferrer-i-Cancho}.
\newblock Compression and the origins of {Zipf's} law for word frequencies.
\newblock {\em Complexity}, 21:409--411, 2016.

\bibitem{Simon}
H.~A. Simon.
\newblock On a class of skew distribution functions.
\newblock {\em Biomet.}, 42:425--440, 1955.

\bibitem{Cattuto}
C.~Cattuto, V.~Loreto, and L.~Pietronero.
\newblock Semiotic dynamics and collaborative tagging.
\newblock {\em Proc. Natl. Acad. Sci. USA}, 104(5):1461--1464, 2007.

\bibitem{Zanette_2005}
D.~Zanette and M.~Montemurro.
\newblock Dynamics of text generation with realistic {Zipf}'s distribution.
\newblock {\em J. Quant. Linguist.}, 12(1):29--40, 2005.

\bibitem{Gerlach_Altmann}
M.~Gerlach and E.~G. Altmann.
\newblock {Stochastic model for the vocabulary growth in natural languages}.
\newblock {\em Phys. Rev. X}, 3:021006, 2013.

\bibitem{Saichev_Sornette_Zipf}
A.~Saichev, Y.~Malevergne, and D.~Sornette.
\newblock {\em Theory of {Zipf}'s Law and of General Power Law Distributions
  with {Gibrat}'s Law of Proportional Growth}.
\newblock Lecture Notes in Economics and Mathematical Systems. Springer Verlag,
  Berlin, 2009.

\bibitem{Tria}
F.~Tria, V.~Loreto, V.~D.~P. Servedio, and S.~H. Strogatz.
\newblock The dynamics of correlated novelties.
\newblock {\em Sci. Rep.}, 4:05890, 2014.

\bibitem{Perkins}
T.~J. Perkins, E.~Foxall, L.~Glass, and R.~Edwards.
\newblock A scaling law for random walks on networks.
\newblock {\em Nature Comm.}, 5:5121, 2014.

\bibitem{Bak_book}
P.~Bak.
\newblock {\em How Nature Works: The Science of Self-Organized Criticality}.
\newblock Copernicus, New York, 1996.

\bibitem{Sethna_nature}
J.~P. Sethna, K.~A. Dahmen, and C.~R. Myers.
\newblock Crackling noise.
\newblock {\em Nature}, 410:242--250, 2001.

\bibitem{Sornette_critical_book}
D.~Sornette.
\newblock {\em Critical Phenomena in Natural Sciences}.
\newblock Springer, Berlin, 2nd edition, 2004.

\bibitem{Watkins_25years}
N.~W. Watkins, G.~Pruessner, S.~C. Chapman, N.~B. Crosby, and H.~J. Jensen.
\newblock 25 years of self-organized criticality: Concepts and controversies.
\newblock {\em Space Sci. Rev.}, 198:3--44, 2016.

\bibitem{Jaynes57}
E.~T. Jaynes.
\newblock Information theory and statistical mechanics.
\newblock {\em Phys. Rev.}, 106:620--630, 1957.

\bibitem{Nieves}
V.~Nieves, J.~Wang, R.~L. Bras, and E.~Wood.
\newblock Maximum entropy distributions of scale-invariant processes.
\newblock {\em Phys. Rev. Lett.}, 105:118701, 2010.

\bibitem{Main_information}
I.~G. Main and P.~W. Burton.
\newblock Information theory and the earthquake frequency-magnitude
  distribution.
\newblock {\em Bull. Seismol. Soc. Am.}, 74(4):1409--1426, 1984.

\bibitem{Peterson_Dill}
J.~Peterson, P.~D. Dixit, and K.~A. Dill.
\newblock A maximum entropy framework for nonexponential distributions.
\newblock {\em Proc. Natl. Acad. Sci. USA}, 110(51):20380--20385, 2013.

\bibitem{Havrda_Charvat}
J.~Havrda and F.~Charv\'at.
\newblock Quantification method of classification processes. concept of
  structural $a$-entropy.
\newblock {\em Kybernetika}, 3:30--35, 1967.

\bibitem{Tsallis_bjp}
C.~Tsallis.
\newblock {Nonextensive statistics: theoretical, experimental and computational
  evidences and connections}.
\newblock {\em {Braz. J. Phys.}}, 29:1--35, 03 1999.

\bibitem{Hanel_Thurner}
R.~Hanel and S.~Thurner.
\newblock A comprehensive classification of complex statistical systems and an
  axiomatic derivation of their entropy and distribution functions.
\newblock {\em Europhys. Lett.}, 93:20006, 2011.

\bibitem{Hanel_Thurner_2}
R.~Hanel and S.~Thurner.
\newblock When do generalized entropies apply? {How} phase space volume
  determines entropy.
\newblock {\em Europhys. Lett.}, 96(5):50003, 2011.

\bibitem{Stephens_Bialek}
G.~J. Stephens and W.~Bialek.
\newblock Statistical mechanics of letters in words.
\newblock {\em Phys. Rev. E}, 81:066119, 2010.

\bibitem{Broderick}
T.~Broderick, M.~Dud\'{\i}k, G.~Tkacik, R.~E. Schapireb, and W.~Bialek.
\newblock Faster solutions of the inverse pairwise {Ising} problem.
\newblock {\em arXiv}, 0712.2437, 2007.

\bibitem{Berger_etal}
A.~L. Berger, S.~A.~D. Pietra, and V.~J.~D. Pietra.
\newblock A maximum entropy approach to natural language processing.
\newblock {\em Comp. Ling.}, 22:39--71, 1996.

\bibitem{Gerlach_Font_Clos}
M.~Gerlach and F.~Font{-}Clos.
\newblock A standardized {Project Gutenberg} corpus for statistical analysis of
  natural language and quantitative linguistics.
\newblock {\em arXiv}, 1812.08092, 2018.

\bibitem{Mandelbrot61}
B.~Mandelbrot.
\newblock {On the theory of word frequencies and on related {Markovian} models
  of discourse}.
\newblock In R.~Jakobson, editor, {\em Structure of Language and its
  Mathematical Aspects}, pages 190--219. American Mathematical Society,
  Providence, RI, 1961.

\bibitem{Corral_Deluca}
A.~Deluca and A.~Corral.
\newblock Fitting and goodness-of-fit test of non-truncated and truncated
  power-law distributions.
\newblock {\em Acta Geophys.}, 61:1351--1394, 2013.

\bibitem{Corral_Gonzalez}
A.~Corral and A.~Gonz\'alez.
\newblock Power law distributions in geoscience revisited.
\newblock {\em Earth Space Sci.}, 6(5):673--697, 2019.

\bibitem{Corral_nuclear}
A.~Corral, F.~Font, and J.~Camacho.
\newblock Non-characteristic half-lives in radioactive decay.
\newblock {\em Phys. Rev. E}, 83:066103, 2011.

\bibitem{Voitalov_krioukov}
I.~{Voitalov}, P.~{van der Hoorn}, R.~{van der Hofstad}, and D.~{Krioukov}.
\newblock {Scale-free Networks Well Done}.
\newblock {\em arXiv}, 1811.02071, 2018.

\bibitem{Corral_Cancho}
A.~Corral, I.~Serra, and R.~Ferrer{-i-}Cancho.
\newblock The distinct flavors of {Zipf}'s law in the rank-size and in the
  size-distribution representations, and its maximum-likelihood fitting.
\newblock {\em {arXiv}}, 1908:01398, 2019.

\bibitem{Corral_csf}
A.~Corral.
\newblock Scaling in the timing of extreme events.
\newblock {\em Chaos. Solit. Fract.}, 74:99--112, 2015.

\bibitem{Burroughs_Tebbens}
S.~M. Burroughs and S.~F. Tebbens.
\newblock Upper-truncated power laws in natural systems.
\newblock {\em Pure Appl. Geophys.}, 158:741--757, 2001.

\bibitem{Berger}
A.~Berger.
\newblock The improved iterative scaling algorithm: A gentle introduction.
\newblock {\em preprint}, 1997.

\bibitem{BHP}
S.~T. Bramwell, K.~Christensen, J.-Y. Fortin, P.~C.~W. Holdsworth, H.~J.
  Jensen, S.~Lise, J.~M. L\'opez, M.~Nicodemi, J.-F. Pinton, and M.~Sellitto.
\newblock Universal fluctuations in correlated systems.
\newblock {\em Phys. Rev. Lett.}, 84:3744--3747, 2000.

\bibitem{Font_clos_molon}
F.~Font-Clos and N.~R. Moloney.
\newblock Percolation on trees as a {Brownian} excursion: from {Gaussian} to
  {Kolmogorov-Smirnov} to exponential statistics.
\newblock {\em Phys. Rev. E}, 94, 2016.

\bibitem{Corral_garcia_moloney_font}
A.~Corral, R.~Garcia-Millan, N.~R. Moloney, and F.~Font-Clos.
\newblock Phase transition, scaling of moments, and order-parameter
  distributions in {Brownian} particles and branching processes with
  finite-size effects.
\newblock {\em Phys. Rev. E}, 97:062156, 2018.

\bibitem{Tria_Loreto_music}
J.~Sakellariou, F.~Tria, V.~Loreto, and F.~Pachet.
\newblock Maximum entropy models capture melodic styles.
\newblock {\em Sci. Rep.}, 7:9172, 2017.

\end{thebibliography}

\end{document}